\documentclass[twocolumn,preprintnumbers,amsmath,amssymb, pr]{revtex4}

\usepackage{epsfig}

\begin{document}

\title{Two Parallel Swendsen-Wang Cluster Algorithms Using Message-Passing Paradigm}
\author{Shizeng Lin\(^{1,2,3}\) and Bo Zheng\(^{1}\)}

\affiliation{ \(^{1}\)Zhejiang Institute of Modern Physics, Zhejiang
University, Hangzhou 310027, P.R. China\\
\(^{2}\)WPI Center for Materials Nanoarchitectonics, National Institute for Materials Science, Tsukuba 305-0047, Japan\\
\(^{3}\)Graduate School of Pure and Applied Sciences, University of
Tsukuba, Tsukuba 305-8571, Japan}
\date{\today}

\begin{abstract}
In this article, we present two different parallel Swendsen-Wang
Cluster(SWC) algorithms using message-passing interface(MPI). One is
based on Master-Slave Parallel Model(MSPM) and the other is based on
Data-Parallel Model(DPM). A speedup of $24$ with $40$ processors and
$16$ with $37$ processors is achieved with the DPM and MSPM
respectively. The speedup of both algorithms at different
temperature and system size is carefully examined both
experimentally and theoretically, and a comparison of their
efficiency is made. In the last section, based on these two parallel
SWC algorithms, two parallel probability changing cluster(PCC)
algorithms are proposed.
\end{abstract}

\maketitle

\section{Introduction}
In the last decades, due to the great improvement in the computer
performance, computer simulations become a more and more powerful
tool in exploring the nature of many physical phenomena\cite{lanbo}.
Because the computer simulations can tune the microscopic details of
the system, they provide a deep insight view of the physical system.
However, because of the limitation of the CPU time and memory space,
computer simulations can only investigate finite systems and thus
all the simulation results carry certain finite-size effect. Though
the finite-size scaling\cite{barbo} analysis can be used to obtain
the infinite-size results by extrapolating the finite-size results,
it is still rather appealing to investigate the system as large as
possible, for the exact scaling function of physics system doesn't
know and the scaling function is invalid for small system size $L$
in many systems. Thus an efficient parallel algorithm is of great
importance and demanding.

It is well-known that near the criticality, the so-called critical
showing down effect will occur because of the divergence of
auto-correlation time $\tau$. This effect is rather harmful to
simulations for one can not obtain independent configurations in the
critical region. During the last two decades, much efforts have
been taken to solve this problem\cite{swe87,zhe98}.
The Swendsen-Wang Cluster(SWC) algorithm is one among the most
successful algorithm which
can largely reduce the dynamic exponent $z$\cite{bai91,wan02,Du06}.
 Taking the Ising model $H=-J\sum_{<ij>}\sigma_i\sigma_j$ as an
example, this non-rejecting, global flipping cluster algorithm
consists following steps:
\begin{enumerate}
\item Traverse all the links connecting neighbor spins $\sigma_i$
and $\sigma_j$. $\sigma_i=\pm1$ is the spin at site $i$. If
$\sigma_i = \sigma_j$, connect these two spins with probability
$P=1-\exp(-2J/T)$ with $T$ being the temperature. Connected links
are also called bonds.
\item Construct the cluster, two spins belong to a cluster if and
only if they are connected by a path of a connected links.
\item Assign a new random spin state to these clusters, all spins which
belong to the same cluster have the same new spin state.
\item Goto the first step.
\end{enumerate}

This cluster algorithm has achieved great success after it was
proposed. Many extensions of this algorithm were
proposed\cite{Wolff89,liu04}. One of them is the Probability
Changing Cluster(PCC) algorithm\cite{Oka01}. By introducing feedback
mechanism into the SWC algorithm, it can tune the critical
temperature automatically. This powerful method shows its merit in
many systems where the SWC algorithm is
applicable\cite{oka012,oka02}. However, unlike the molecular
dynamics simulation which is inherent parallelism, the SWC needs
extra technics to parallelize it. There are many works on
parallelizing the SWC algorithm for specific
machines\cite{heer90,fla92,ker92,bar94} and these methods have
attained scalability with different degrees of success. There exist
two commonly used parallel algorithm models for SWC algorithm: MSPM
and DPM. Taking the Ising model as an example again, in the former
model, the lattice is decomposed into domains and then each domain
is assigned to a processor called slave. All slave processors
construct the cluster independently. When the slave processors
finish constructing local clusters, another processor called master
processor collects all the border clusters, connects them to form
global clusters and broadcasts all the global clusters to all slave
processors. In the latter model, the lattice is again decomposed
into domains and each domain is assigned to a processor. Each
processor connects spin simultaneously and sends its border cluster
to neighbor processors. When neighbor processor receives the border
cluster, synchronization is made to form global clusters. Iterating
the communications until all global clusters are constructed. The
detailed parallel algorithm of both models will be described in the
following section.

In this article, we improved the algorithm in Ref. \cite{bar94} and
present a new DPM parallel algorithm which is similar to Ref.
\cite{fla92}. We analyze speedup of these algorithms with different
temperature and different system size on the Dawning 4000A and SGI
Onyx 3900 supercomputer. In the last section, we present two
parallel PCC algorithms based on these two parallel algorithms. The
remaining part of this article is organized as: In the Sec. II (A)
and (B), we describe both the MSPM and DPM algorithms separately. In
the Sec. II (C), we compare the performance of these two algorithms.
In the Sec. III, a parallel program for PCC is proposed. Then comes
the conclusion.

\section{Two Parallel SW Algorithms}
\subsection{Master-Slave Parallel Model}\label{mspm}
\begin{figure}[b]
\psfig{figure=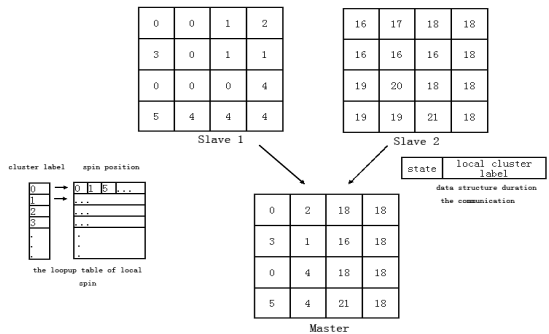,width=\columnwidth} \caption{Illustration of
the MSPM. The digits in the squire lattice denote the label of the
local cluster. The leftmost is the lookup table of cluster, the
entry is the label of the cluster and it points to the address of
the spin which belongs to this cluster. The rightmost is the data
structure during the gathering and broadcasting
communication.}\label{ms}
\end{figure}

In the MSPM, the lattice is divided into domains and the domains are
assigned to slave processors. The data decomposition technics is
very important here since the bad data decomposition will cause the
master processor overloaded\cite{grabo}. We will extend this point
later. Clusters are classified into two types: local clusters in
which the whole cluster is in one domain, and global clusters in
which the cluster spans more than one domain. Each slave processor
is responsible for constructing clusters inside the domain and
assigning a new state to them. However, conflicts arise when global
cluster emerges. In this parallel model, a master processor is
introduced to solve the conflicts. For clusters who have border
sites are the candidate of global clusters, the master processor
gathers all the border sites, creates "new lattice" with these
border sites, constructs clusters and assigns a new state to the
"new lattice". When the master processor finishes these steps, it
scatters all the global clusters to the slave processors. The slave
processors then flip the local clusters using the received
information. This is illustrated in the Fig. \ref{ms}.

In this parallel model, the slave processors for one Monte Carlo
Sweep(MCS) have the following steps:
\begin{enumerate}
\item Construct clusters using Hoshen-Kopelman cluster counting
algorithm\cite{hos76}. Each cluster has a unique identity through
the domains. Meanwhile, a cluster lookup table is built up which is
intended to improve the efficiency of traversing all sites belonging
to a cluster. As shown in Fig. \ref{ms}, the entry of this table is
the identity of the cluster and the nodes are the pointers pointing
to all sites belonging to this cluster. Clusters in this table are
marked with border flag if they have border sites. Otherwise they
are marked with interior flag. Each site also has a pointer which
points to the entry of the lookup table. With the help of these data
structures, it is very convenient to find all sites belonging to a
cluster and to find the cluster to which this site
belongs\label{hk}.

\item During this communication step, the slave processors send
all the border sites to the master processor. For the illustration
as described in Fig. \ref{ms}, we take the column border as an
example. Each slave processor sends its leftmost and rightmost
columns to the master processor. The communication information
contains the identity of the cluster to which the sites belong and
state of the sites.

\item Traverse all the sites which belong to local clusters marked with
interior flag and assign a random spin state to them. Now the sites
residing in the same cluster have the same random spin state.

\item When the master processor finish constructing the global
clusters, the slave processors receive the global cluster data
structure from the master processor. It has the same data structure
as the sending data. Each slave processor visits all sites residing
in the same cluster from the entry of the lookup table and assigns
the received new state to these sites.
\end{enumerate}

The master processor for one MCS consists the following steps.
\begin{enumerate}
\item Construct the "new lattice" with the received sites. Here we take
the column sites as an example. The master processor aligns the
received leftmost and rightmost columns from the left domain to the
right domain. This step is shown in Fig. \ref{ms}.
\item Construct the clusters using Hoshen-Kopelman cluster counting
algorithm and assign a new state to this cluster. There is a little
difference between the procedure done here and the procedure done in
the slave processors. For the columns who are originally in the same
slave processors, they have already connected to form clusters in
the step \ref{hk}, one needn't to activate these link again. One
only needs to activate the link between neighbor domains. To make
the situation clearly, we take Fig. \ref{ms} as an example. Because
the spins in the first column and second column have already be
visited to form clusters, we only need to activate the link between
the second column and the third column. After finishing this step,
each cluster now has a global cluster label.

\item Scatter all the cluster identity and its new state to all slave processors.
\end{enumerate}

The advantages of this algorithm are as follows. For ease of
discussion, a two dimensional lattice is considered throughout this
paper. Firstly, in our implement of this algorithm, the slave
processors are mapped into two dimensional square array to minimize
the communication. let $N$ denote the number of slave processors and
$L$ denote the linear size of the lattice. In this division
scenario, each slave processor has domain of lattice whose size is
$L^2/N$. On the other hand, the size of border sites received from
slave processors by the master processor is $4L\sqrt N$. Since the
inequality $L>4N\sqrt N$ always holds in the simulation, this
division method avoid overload in the master processor. To some
extent, this division method can remove the bottleneck mentioned in
Ref. \cite{bar94}.

Secondly, the algorithm in Ref. \cite{bar94} used a global memory,
thus the communication costs is negligible in that case. However, in
modern supercomputers, the MPI is always used to exchange the
message between different processors. In this paper, we intend to
devise a new algorithm based on MPI. For the communication is more
time-consuming than the computation, the performance of this
algorithm critically relies on the performance of the communication.
We will give experimental results about this point in the Sec.II
(C).

The time cost of this algorithm can be counted as follows. Let $L^2$
denote the total number of sites, $N$ is the number of slave
processors, $\tau_w$ is the per-integer transfer time,
$\tau_{\rm{site}}$ is the per-site bond-activating time, and $t_i$
is the time cost in the $i$th step mentioned above. The domain size
in each processor is $L^2/N$. For simplicity, we assume the lattice
is divided into square array of domains. Let us define the time
required by slave processors and mater processor for step $i$ by
$t_{si}$ and $t_{mi}$ respectively. The time cost in the first step
of the slave processors is
\begin{equation}
t_{s1}=\frac{L^2}{N} \times \tau_{\rm{site}}.
\end{equation}
The second step of slave processors is that the master processor
gathers all boundary sites from each domains. The time cost in the
gathering process is
\begin{equation}
t_{s2}=\frac{4L\tau_w}{\sqrt N}\times (N-1)\approx 4L \tau_w \sqrt
N,
\end{equation}
where we have omitted the startup time $t_s$ for communication.
Since the third step of slave processors and the first and second
steps of master processor are executed simultaneously, the time cost
is the maximum time needed for these steps. We have used a linear
algorithm to create "new lattice", therefore the time cost of this
step scales linearly with the size of the "new lattice". We denote
$\tau_{m1}$ the time cost for one site. The total size of this "new
lattice" is $4L \sqrt N$. Thus we have
\begin{equation}
t_{m1}=4L \sqrt N \tau_{m1}.
\end{equation}
The time cost in the second step of master processor is the same as
the first step in slave processors except for the size of lattice.
Therefore we have
\begin{equation}
t_{m2}=2L \sqrt N \tau_{\rm{site}}.
\end{equation}

The time cost in the third step of master processor depends on the
number of global clusters, therefore it depends on the simulation
temperature. Let $C_{\rm{total num}}$ denote the total number of the
global clusters. The time cost in this scattering process is
\begin{equation}
t_{m3}=\frac{C_{\rm{total num}}}{N} \tau_w (N-1)\approx C_{\rm{total
num}} \tau_w.
\end{equation}

With the help of the lookup table, the time cost in fourth step of
the slave processors depends on the number of clusters received by
each slave processors. For simplicity, we use average size of
cluster in each slave processor.
\begin{equation}
t_{s4}=\frac{C_{\rm{total num}}}{N} \tau_{s4}.
\end{equation}
The total time cost of this algorithm in one MCS therefore is
\begin{widetext}
\begin{eqnarray}\label{mseq}
t_{\rm{total}}=t_{s1}+t_{s2}+\max(t_{s3},t_{m1}+t_{m2})+t_{m3}+t_{s4} \\
=\frac{L^2\tau_{\rm{site}}}{N}+8L \tau_w \sqrt N+\max(t_{s3},4L\sqrt
N \tau_{m1}+2L \sqrt N \tau_{\rm{site}})+C_{\rm{total num}}
\tau_w+\frac{C_{\rm{total num}}}{N} \tau_{s4} \\
\equiv \frac{L^2\tau_{\rm{site}}+C_{\rm{total num}} \tau_{s4}}{N} +L
\sqrt N \tau_{\rm{eff}}+ \rm{constant}
\end{eqnarray}
\end{widetext}
Equation (\ref{mseq}) is a little complicate, however, as will be
shown in the computer experiments, many terms are negligibly small
and this equation will reduce to a single form. This will be shown
later.

\subsection{Data Parallel Model}\label{dpm}
As mentioned in the Sec. II(A), conflicts arise when processor flips
the spins which belong to a global cluster. In the MSPM, the
conflicts are solved by introducing a master processor. However, in
the DPM, the conflicts are solved by the communication between all
processors. The lattice is again divided into domains and assigned
to a processor in the DPM.

\begin{figure}[b]
\psfig{figure=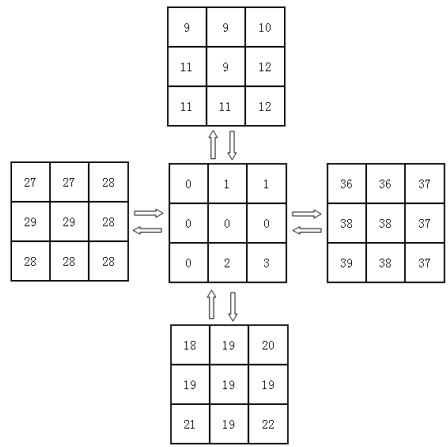,width=\columnwidth } \caption{Illustration of
the DPM. The digits in the squire lattice denote the label of the
local cluster. The arrows in the figure denote that each processor
communicates with its four nearest neighbor processors.}\label{db}
\end{figure}

For the DPM, each processor for one MCS consists the following
steps:
\begin{enumerate}
\item Visit all sites inside the domain and active the link between
two nearest-neighbor sites with probability $P=1-\exp(-2J/T)$ if
they have the same spin state. This procedure can be done using
Hoshen-Kopelman cluster counting algorithm. Each cluster must have a
unique identity. As in the MSPM, a lookup table is also built up and
all sites have a pointer pointing to the cluster which they belong
to. \label{dpstep1}

\item Interchange the state of its boundary sites to the nearest-neighbor
processors in each direction. A reference array is built up to mark
whether the two aligned neighbor sites belong to the same cluster or
not. This step is illustrated in Fig. \ref{db}.

\item For each cluster residing in this domain, pick a random spin sate
and assign it to this cluster while the state of the sites in this
cluster is not changed. In this step, the conflicts between
different processors are disregarded.

\item \label{dpm3} For each processor, interchange the boundary labels and
the new spin state of the boundary sites with the nearest-neighbor
processors in each direction.

\item \label{scf} If the received sites are in the same cluster with the aligned sites, the processor
compares the local label with the received label. If the received
label is less than the local one, assign the received label and the
received new spin state to this cluster.

\item Goto step \ref{dpm3} until no change is found in the step
\ref{scf}. With the help of the lookup table, assign the new spin
state to all sites inside this clusters. Now, all conflicts are
solved and a MCS is finished.

\end{enumerate}

At first sight, this parallel algorithm is very time-consuming for
many communications are involved before all global clusters are
constructed. However, unlike the MSPM, the performance depends on
the number of global clusters, i.e. the simulation temperature, the
communication data is constant in the DPM. Thus the performance of
this parallel model is temperature-independent.

To count the time cost, we use the same notations introduced above.
For simplicity, we also assume the lattice is divided into square
array of domains. The time cost in the first step is
\begin{equation}
t_1=\frac{L^2}{N} \times \tau_{\rm{site}}.
\end{equation}
In the step $2$, the processor needs to interchange its boundary
spin state with its nearest neighbor processors. Thus the time cost
at this step is
\begin{equation}
t_2=\frac{4L}{\sqrt N} \tau_w .
\end{equation}
The factor $4$ denotes four directions in the case of two
dimensional system. In the communication step $4$, the processor
needs to send its boundary label to the furthest processors in the
worst situation when a global cluster spans the who lattice. This
happens near the critical region where the cluster percolates. In
this case, it needs $\sqrt N/2$ iterations in the step $4$. The time
cost of this step is
\begin{equation}
t_4=4\frac{\sqrt N}{2} \times \frac{2L}{\sqrt N} \tau_w=4L\tau_w.
\end{equation}
The factor $4$ also accounts for the four directions. Because the
time cost in step $3,5,6$ is less expensive and is proportional to
$1/N$, we write the total time costs in steps $3,5,6$ as
\begin{equation}
t_{356}=\frac{\tau_{356}}{N}.
\end{equation}
Thus the total time cost $t_{\rm{total}}$ in one MCS is
\begin{widetext}
\begin{equation}\label{dpeq}
t_{\rm{total}}=\sum t_i=\frac{L^2}{N} \tau_{\rm{site}}+
\frac{4L}{\sqrt N} \tau_w +t_3+ \frac{\sqrt N}{2} \times
\frac{8L}{\sqrt N} \tau_w +t_5 +t_6=4(1+\frac{1}{\sqrt N}) L \tau_w+
\frac{L^2 \tau_{\rm{site}}+\tau_{356}}{N}
\end{equation}
\end{widetext}
In the implement of this algorithm, we found that $4(1+1/\sqrt N) L
\tau_w \ll (L^2 \tau_{\rm{site}}+\tau_{356})/{N}$, thus the speedup
of the DPM scales linearly with the number of processors.

\subsection{Performance and Comparison}\label{percom}
\begin{figure}[b]
\psfig{figure=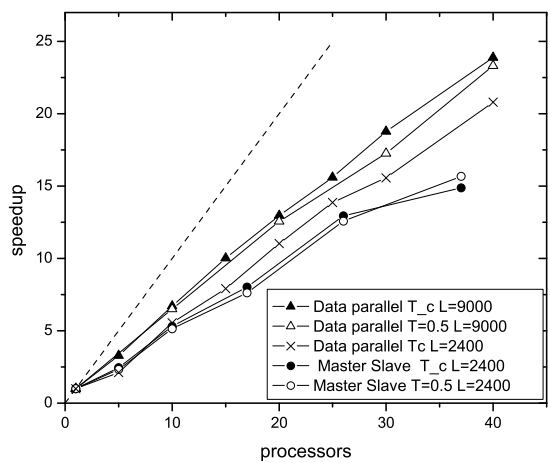,width=\columnwidth} \caption{Speedup of
the MSPM and the DPM at different temperature and different system
size on the SGI Onyx 3900 supercomputer. The dotted line is the
ideal speedup who has a slope of one.}\label{speedup}
\end{figure}

\begin{figure}[b]
\psfig{figure=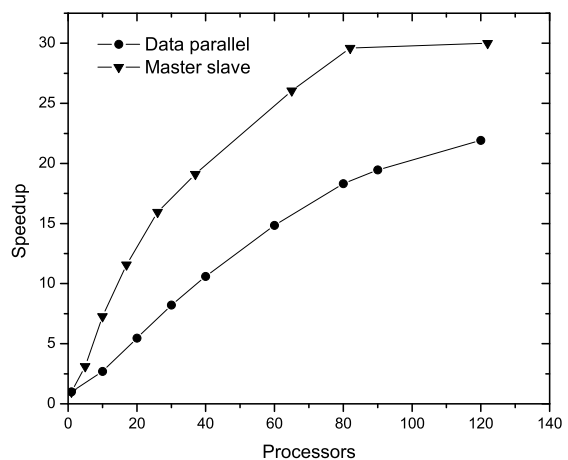,width=\columnwidth} \caption{Speedup of
the MSPM and the DPM on the Dawning 4000A supercomputer. The linear
size of the simulated system is $L=7920$ and temperature is
$T=T_c$.}\label{dawning}
\end{figure}

We analyze the performance of these two algorithms on the Onyx3900
supercomputer and Dawning 4000A supercomputer. The results are
compiled in Fig. \ref{speedup} and Fig. \ref{dawning}. The results
in these figures give the speedup of the two algorithms with
different lattice size and temperature. Here are a few observations
which can be made when analyzing these figures.

\begin{itemize}
  \item A first observation is that both two algorithms are
  independent of the temperature, i.e. the number of total global clusters. It
  is reasonable in DPM for the scaling function Eq.
  (\ref{dpeq}) is independent of the number of the global clusters. However,
  in the MSPM, $t_{\rm{total}}$ in Eq. (\ref{mseq}) explicitly depends on
  the number of global clusters. We measured the time cost of different steps in
  Eq. (\ref{mseq}) and found that this term is not dominant.

  \item The MSPM in this paper has a linear
  speedup. In Ref. \cite{bar94}, the author found that their
  algorithm only has the maximum speedup of $3$, as the number of processors
  increases. This is due to the overload of the master processor. In this
  paper, by appropriate decomposition and message passing, the
  master processor never gets overloaded in this case. Thus, the MSPM algorithm presented in this paper to some extent
  improved the algorithm in Ref. \cite{bar94}.

  \item As the $L$ increases, the speedup of the DPM will
  also increase. At first sight this is unreasonable since the increase
  in the lattice size will reduce the cache hit rate and therefore
  reduce the performance of this algorithm. Nevertheless, the
  speedup of the DPM is $S_{\rm{dp}}\sim {L^2}/({aL+bL^2})$.
  It is straightforward to see that as the $L$ increases, the speedup
  will also increases.

  \item The speedup of the MSPM exceeds the speedup of the DPM on the Dawning 4000A supercomputer while on the SGI
  Onyx3900 the situation reverses. This is due to the difference in the architecture
   of these two supercomputers. The communication between the
   different nodes on the Dawning supercomputer uses the second level network connection
   which costs more time than that of the SGI onyx3900 supercomputer.
   Because the DPM involves massive and many steps
   communication, the performance deteriorates on slow network. On
   the other hand, when the performance of the communication is largely enhanced,
   the speedup of the DPM will exceed that of the
   MSPM as in the case of the SGI onyx3900 supercomputer
    for the MSPM still carries many serial
   elements. The master/slave processor needs to wait before slave/master processor finishes.
\end{itemize}

\section{Parallel PCC Algorithm}\label{ppa}
The basic notion in the PCC algorithm is the criteria of
percolation. In finite-size systems, there are various criteria of
percolation and each of them may lead to different results. However,
in the thermodynamic limit, the results obtained by different
criteria converge. Two criteria are presented in Ref. \cite{Oka01},
one of which is the extension rule and the other is the topological
rule. The former rule is that one cluster spans the whole simulation
box at least one of the $d$ directions in d-dimensional systems. The
latter rule is that one cluster winds around the simulation box at
least one of the $d$ directions in d-dimensional systems. One can
also devise other criteria if and only if they will converge in the
thermodynamic limit.

\begin{figure}[t]
\psfig{figure=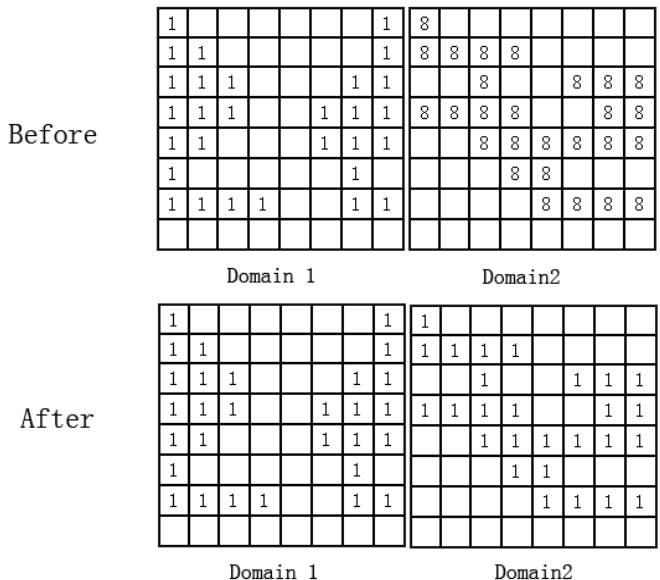,width=\columnwidth} \caption{Illustration of
the calculation the range of a cluster. The digit number is the
label of a cluster and domain 1 and 2 are adjoint domains. The top
part of this figure shows the cluster before activating links
between domain 1 and 2 while the bottom part is the cluster after
activating links.}\label{percolating}
\end{figure}

In our study, we take extension rule as an example. It is very
similar with different criteria in detailed algorithm. A naive
implement of the extension rule is by enumerating pathes of all
clusters and counting whether there exists a cluster spanning the
whole lattice. However this simple method is inefficient, so we use
another way. For the open boundary condition, it is straightforward
to count the spanned cluster in that one only needs to count whether
there is a cluster occupying both leftmost and topmost boundary or
topmost and bottommost boundary. For the periodic boundary
condition, because there is no topmost and bottommost boundary, the
situation is a little complicate. One has to project all clusters
onto x axis or y axis and calculate the length of these clusters,
this is illustrated in Fig. \ref{percolating}.
 let $P_i(C)$ is a range operator which calculate
the range of cluster $C$ in \emph{i}th dimension. Before activating
the links between domain 1 and domain 2, if we only consider
$P_x(C)$, then, $P_x(1)=[0,3]\bigcup [5,7]$ and $P_x(8)=[8,15]$.
After activating, the cluster 1 in domain 1 and cluster 8 in domain2
are merged, the ranges of these clusters must be updated, i.e.
$P_x(1)=[0,3]\bigcup[5,15]$. After all global clusters are
constructed, it is easy to find out the spanned cluster. In Fig.
\ref{percolating}, the system is not percolating for the range of
the global cluster 1 doesn't cover the whole range of the system.

For the open boundary condition case, it is straightforward to count
the spanned cluster, we only concentrate our attention on the
periodic boundary condition which is widely used in computer
simulations. In the Step 1 of both the MSPM and the DPM, range of
projected cluster is calculated. The range of a cluster, together
with its label is gathered by master processor in the MSPM and is
sent to its neighbor processors in the DPM. When clusters are merged
into larger cluster, the range of merged clusters must be updated.
When all global clusters are constructed, in the MSPM, the master
processor enumerates all global clusters to find out whether there
is a cluster whose range is $[0,L]$ in one direction, while in the
DPM, all processors traverse all clusters inside its domain to find
out whether there exists a percolating cluster, because the range of
the clusters inside each domains are the range of global cluster at
this domain.

In these two parallel PCC algorithms, extra time is only needed at
the first step and final step in both MSPM and DPM. We have measured
the speedup of these two parallel PCC algorithms and find the time
cost is very similar to that of MSPM and DPM.

\section{conclusion}
We present two parallel Swendsen-Wang Cluster algorithms based on
Master-Slave Parallel Model and Data-Parallel Model using
message-passing interface. The speedup of these two parallel
algorithms is measured both on the SGI Onyx3900 and Dawning 4000A
supercomputers. The scaling function of the time cost for both
algorithms are derived. At last section, two parallel probability
changing cluster algorithms are proposed.

\section{Acknowledgement}
The authors would like to thank R. Ziff for helpful discussion with
the percolation criteria during the 2nd International Workshop on
Computational Physics, Hangzhou. They are also indebt to Prof. J.
-S. Wang at National University of Singapore for hospitality where
partial of this work was done. The simulations are carried out at
Center for Engineering and Scientific Computation, Zhejiang
University and Shanghai Supercomputer Center.


\end{document}